\documentclass[mathleft
]{an}

\usepackage{graphicx}
\usepackage{times}
\usepackage{natbib}
\bibpunct{(}{)}{;}{a}{}{,}

\begin{document}

\Pagespan{1}{}
\Yearpublication{2010}%
\Yearsubmission{2010}%
\Month{11}%
\Volume{999}%
\Issue{88}%

\title{Mode transformation and frequency change with height in 3D numerical simulations of magneto-acoustic wave propagation in sunspots}

\author{T. Felipe\inst{1,2}\fnmsep\thanks{Corresponding author:
  \email{tobias@iac.es}\newline}
\and  E. Khomenko\inst{1,2,3}
\and  M. Collados\inst{1,2}
}
\titlerunning{Waves in sunspots}
\authorrunning{T. Felipe \& E. Khomenko \& M. Collados}
\institute{
Instituto de Astrof\'{\i}sica de Canarias, 38205,
C/ V\'{\i}a L{\'a}ctea, s/n, La Laguna, Tenerife, Spain
\and 
Departamento de Astrof\'{\i}sica, Universidad de La Laguna, 38205, La Laguna, Tenerife, Spain
\and 
Main Astronomical Observatory, NAS, 03680, Kyiv, Ukraine}



\abstract{%
  Three-dimensional numerical simulations of magnetoacoustic wave propagation are performed in a sunspot atmosphere with a computational domain covering from the photosphere to the chromosphere. The wave source, with properties resembling the solar spectrum, is located at different distances from the axis of the sunspot for each simulation. These results are compared with the theory of mode transformation and also with observational features. Simulations show that the dominant oscillation frequency in the chromosphere decreases with the radial distance from the sunspot axis. The energy flux of the different wave modes involved, including de Alfv\'en mode, is evaluated and discussed.}

\maketitle

\section{Introduction}
Observations of waves in sunspots show a variety of behaviours depending on the height and the region observed. In the umbral photosphere their power spectrum has a peak at 3.3 mHz (\hbox{period} of 5 min) and their
\hbox{amplitudes} are a\-round a hundred m s$^{-1}$. The frequency of the power peak varies with the height, and at the chromosphere the oscillatory pattern is dominated by waves in the 5-6 mHz band (period of 3 min) with amplitudes of several kilometers per second which \hbox{present} a saw-tooth profile. Several \hbox{mechanisms} have been proposed to explain the chromospheric three min\-ute oscillations: a resonant chromospheric cavity \citep{Zhugzhda+etal1985}; non-linear interaction of pho\-to\-spheric
modes \citep{Gurman+Leibacher1984}; and slow magneto-acoustic \hbox{mode} \hbox{field} \hbox{aligned} propagation from the photosphere to the chromosphere
in the 5-6 mHz band \citep{Centeno+etal2006}.

The interpretation of the observed oscillations in terms of MHD waves possess several difficulties. The velocity of propagation of the magneto-acoustic modes depends on the acoustic velocity ($c_S$) and Alfv\'en velocity ($v_A$), both stratified with height. The magnetized atmosphere of sunspots changes from being gas dominated in the photosphere and below, ($cs>v_A$) to field dominated at the high chromosphere ($c_S<v_A$). Around the layer where $c_S=v_A$ the phase speeds of all modes are similar and different waves can interact, which produces the mode trans\-for\-ma\-tion \citep{Bogdan+etal2003, Cally2006, Khomenko+Collados2006}. The direction and efficiency of the
mode transformation depends on the frequency of the wave and the
angle between the wave vector and the magnetic field
\citep{Cally2006}. 

Most of the simulations of mode refraction,  transformation, etc. have been developed for waves with high frequencies, $i.e.$ above the cut-off frequency of the solar atmosphere, and using 2D atmospheres. We aim to study these phenomena through numerical simulations of waves with the periods observed and using realistic atmospheres of sun\-spots in a 3D case. One of the most important benefits of 3D simulations of wave propagation in magnetized atmospheres is the appearance of the Alfv\'en mode. The work by \citet{Felipe+etal2010} points out that the conversion from the slow to the Alfv\'en mode is effective when the wave vector forms a certain angle different from zero with the magnetic field. Conversion to Alfv\'en waves only was produced when the driver was located out of the axis of the sunspot, where the magnetic field has some inclination. In this contribution, we explore further the conclusions of that work to study waves with 3-5 min periodicity found in sunspots, and analyze several simulations with the driver located at different distances from the center of the sunspot.

\begin{figure*}
\includegraphics[width=165mm]{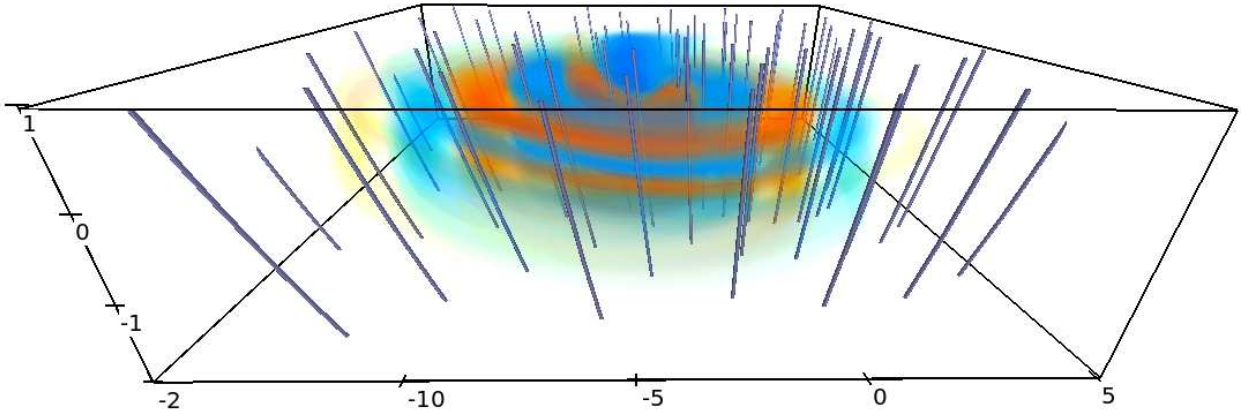}
\caption{Variations of longitudinal velocity scaled by a factor $\rho^{1/2}$ at an elapsed time t=820 s after the beginning of the simulations with the driver
force at 5 Mm from the axis of the sunspot. Grey inclined lines
are magnetic field lines.}
\label{fig:3D_long}
\end{figure*}

\begin{figure*}
\includegraphics[width=165mm]{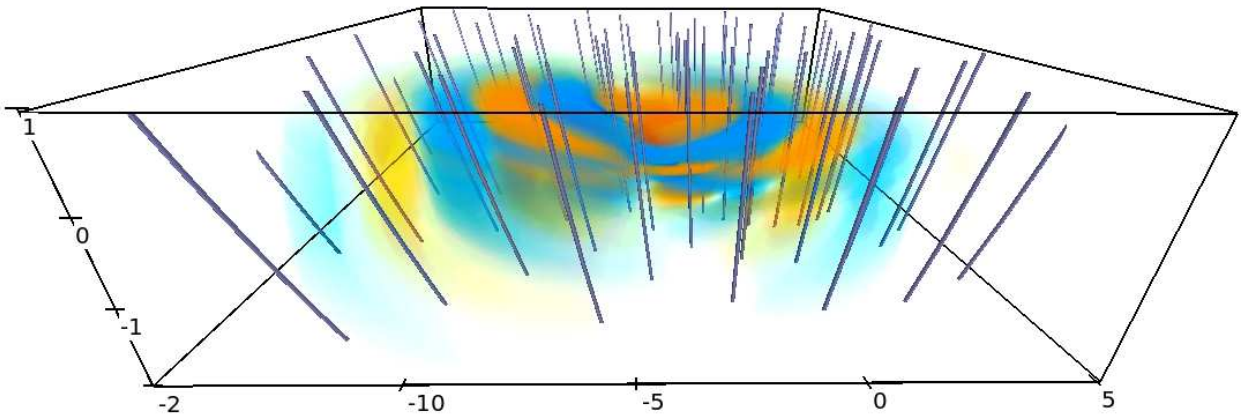}
\caption{Variations of transversal velocity scaled by a factor $\rho^{1/2}$ at an elapsed time t=820 s after the beginning of the simulations with the driver
force at 5 Mm from the axis of the sunspot. Grey inclined lines
are magnetic field lines.}
\label{fig:3D_trans}
\end{figure*}

\section{Numerical simulations}

We have developed a 3D non-linear MHD code \citep{Felipe+etal2010} to study the wave interaction with magnetic structures. We excite a MHS model of sunspot, constructed following \citet{Khomenko+Collados2008}, with a perturbation which drives a wave spectrum close to the solar one. The computational domain covers 60 Mm in both horizontal dimensions, with the axis of the sunspot located at the center, and spanning from $z=-2.5$ to $z=1$ Mm. The height $z= 0$ Mm corresponds to the layer where $\tau_{500}=$1 at quiet Sun photosphere.  The full set of simulations consists of 3 runs with the driver located at $X_0= 5$, $10$ and $15$ Mm from the axis, respectively, and lasting 30 min of solar time. 

Figures \ref{fig:3D_long} and \ref{fig:3D_trans} show the longitudinal and transversal components of the velocity, respectively, of the simulation with the driver located at 5 Mm from the axis of the sunspot. The driver mainly excites a fast acoustic mode below the $\beta=1$ layer. When it reaches this layer it is transformed into a slow acoustic mode (visible in blue in high layers at $x=-5$ Mm in the longitudinal velocity, Fig. \ref{fig:3D_long}), which propagates upwards to the chromosphere; and a fast magnetic mode (visible in the transversal velocity, Fig. \ref{fig:3D_trans}), which is reflected back to the photosphere due to the gradients of the Alfv\'en speed. When it reaches the $\beta=1$ layer, it is transformed again into a fast acoustic mode and a slow magnetic mode in the high-$\beta$ region, which appears in Figs. \ref{fig:3D_long} and \ref{fig:3D_trans}, respectively, as concentric rings around and below $z=0$ Mm. There is also conversion to the Alfv\'en mode, although it is not visible in these figures. We will discuss it in Section \ref{sect:energy}.

\section{Frequency change with height}

\begin{figure}
\includegraphics[width=80mm]{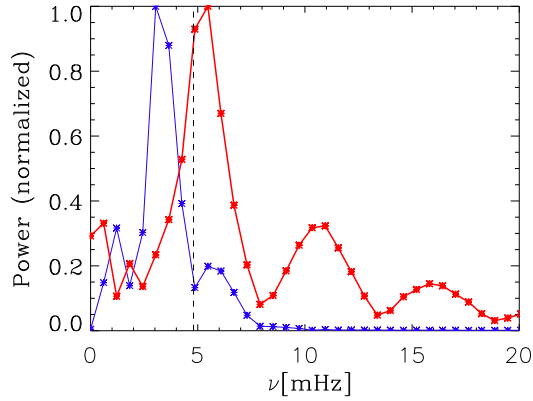}
\caption{Normalized power spectra of the longitudinal velocity at the photosphere (\emph{blue line}) and the chromosphere (\emph{red line}) at the location of the driver for the simulation with $X_0= 10$ Mm. Asterisks mark the measured values. The vertical dashed line marks the maximum value of the cut-off frequency at this position.}
\label{fig:powerspectra}
\end{figure}

Figure \ref{fig:powerspectra} shows the power spectra of the longitudinal velocity at two different heights at the location of the driver for the simulation with $X_0= 10$ Mm. The blue line corresponds to the photospheric height $z= -1$ Mm, where the driver was imposed, while the red line is the power spectra at $z= 1$ Mm, $i.e.$, at the chromosphere. The photospheric power spectra peaks at 3.3 mHz (5 min) and it has a secondary peak around 5.4 mHz. The maximum contribution to the power at the chromosphere is at 5.3 mHz (3 min), so the simulation reflects the observed change of frequency with height.

The power spectra at the chromosphere in Fig. \ref{fig:powerspectra} shows some secondary peaks, which correspond to the harmonics of the 3 min signal, due to the importance of the non-linearities at these high layers, where the longitudinal velocity reaches an amplitude of 4.5 km s$^{-1}$. Waves with frequecies below the cut-off (dashed vertical line) form evanescent waves, which can not propagate energy  upwards, while waves with frequencies above the cut-off (as the secondary peak of the photospheric power spectra) do propagate energy upwards. Their amplitude increases according to the drop of the density.

\begin{figure}
\includegraphics[width=80mm]{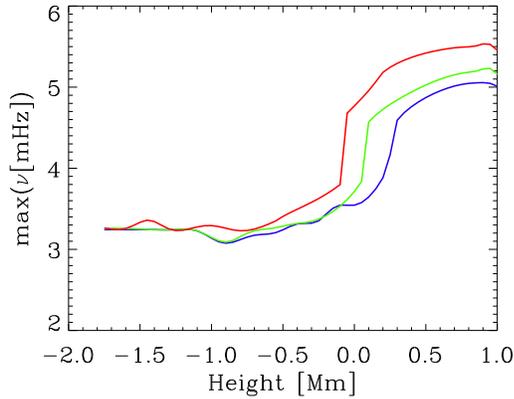}
\caption{Height variation of the dominant frequency of oscillations for the 3 simulations with the driver located at 5 (\emph{red line}), 10 (\emph{green line}) and 15 (\emph{blue line}) Mm of the sunspot axis.}
\label{fig:peakvsheight}
\end{figure}

Figure \ref{fig:peakvsheight} shows the frequency of the dominant power peak at all atmospheric heights. The frequency of the peak at the chromosphere decreases with the radial distance, since the cut-off frequency is higher near the center of the sun\-spot. At photospheric deeper layers the dominant frequency is a\-round 3.3 mHz (5 min) for all simulations with different locations of the driver, increasing abruptly around $z= 0$ Mm (depending on the case), when the power at frequencies above the cut-off becomes more important than the evanescent 3 mHz oscillations.

\section{Energy balance}
\label{sect:energy}
\begin{figure}
\includegraphics[width=80mm]{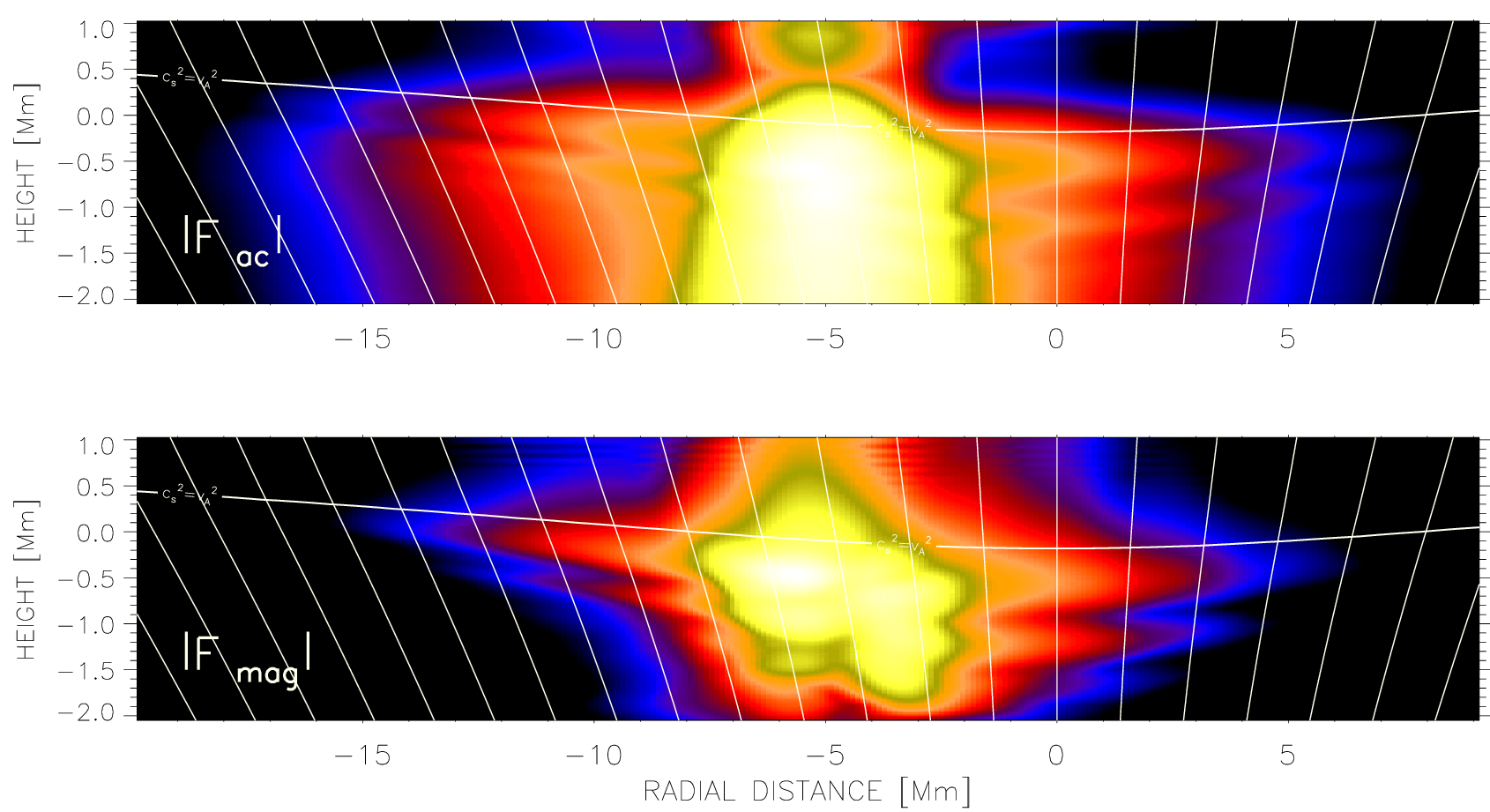}
\caption{Acoustic (\emph{top}) and magnetic (\emph{bottom}) flux for the simulation with $X_0=5$ Mm averaged over the stationary stage of the simulations in logaritmic scale. Horizontal white line is the height where sound velocity and Alfv\'en velocity are equal.}
\label{fig:flux5}
\end{figure}

\begin{figure}
\includegraphics[width=80mm]{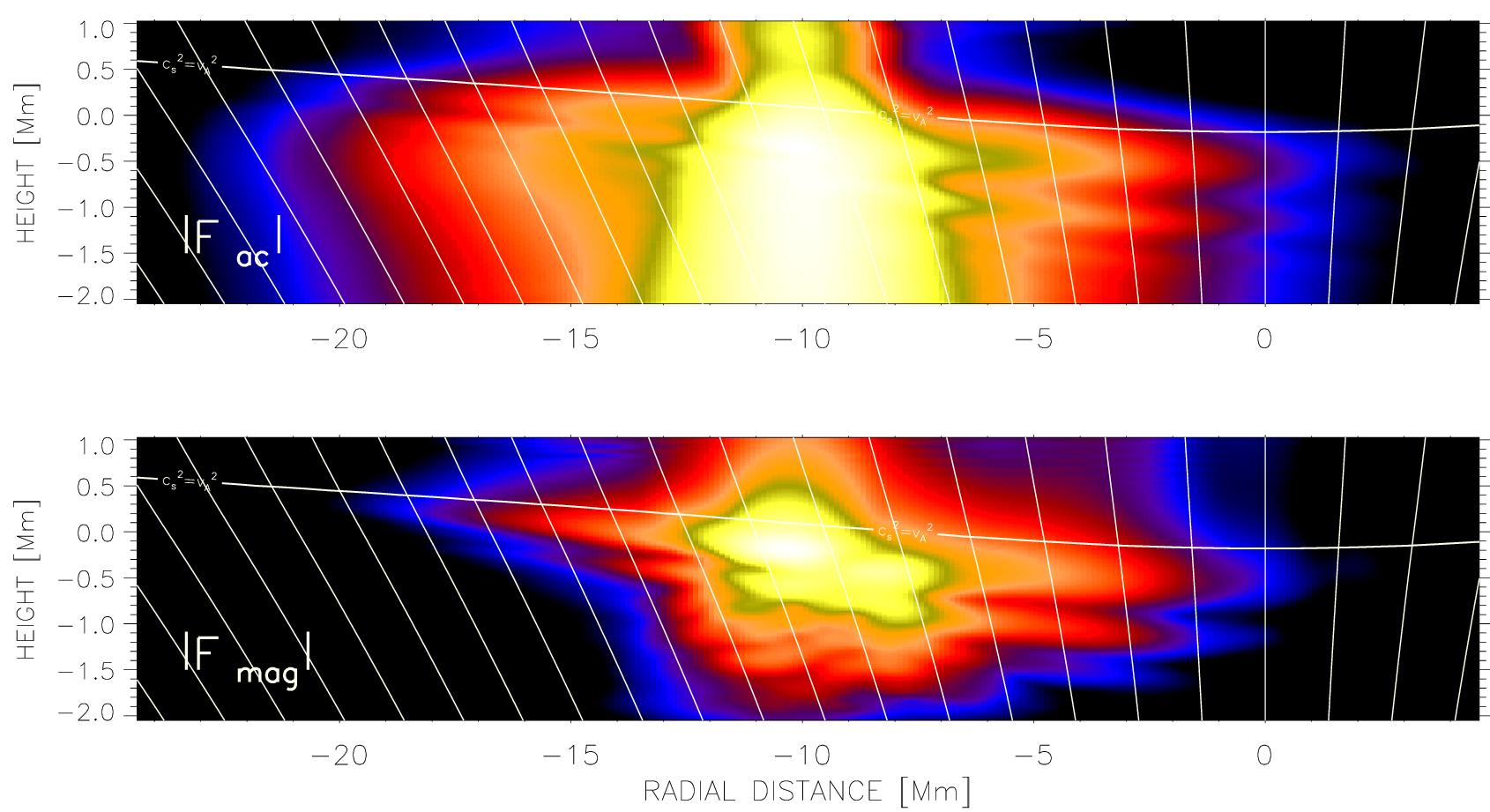}
\caption{Acoustic (\emph{top}) and magnetic (\emph{bottom}) flux for the simulation with $X_0=10$ Mm averaged over the stationary stage of the simulations in logaritmic scale. Horizontal white line is the height where sound velocity and Alfv\'en velocity are equal.}
\label{fig:flux10}
\end{figure}

\begin{figure}
\includegraphics[width=80mm]{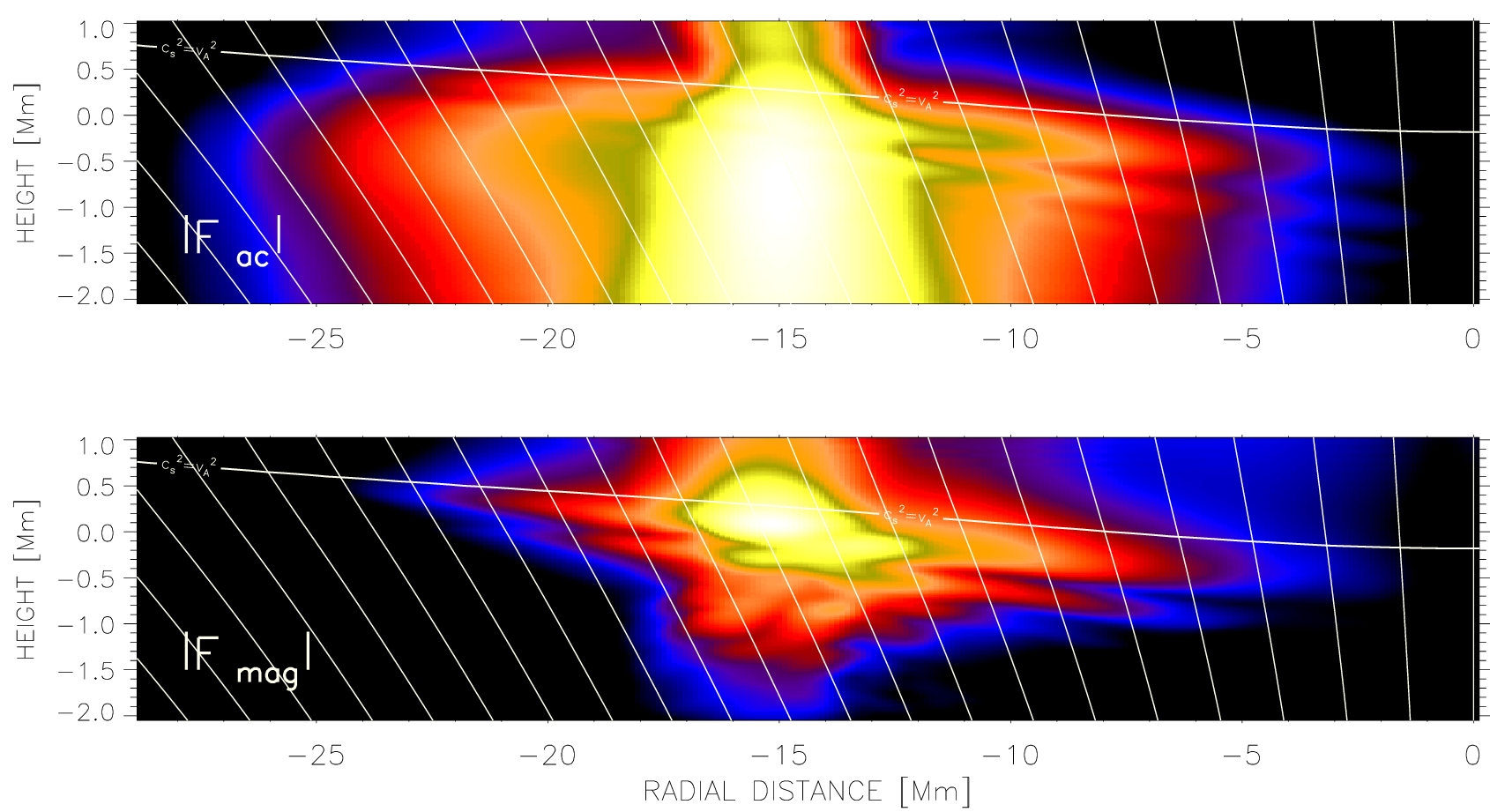}
\caption{Acoustic (\emph{top}) and magnetic (\emph{bottom}) flux for the simulation with $X_0=15$ Mm averaged over the stationary stage of the simulations in logaritmic scale. Horizontal white line is the height where sound velocity and Alfv\'en velocity are equal.}
\label{fig:flux15}
\end{figure}

The driver that we introduce as a perturbation mainly excites a fast acoustic wave in the region where sound speed is higher than Alfv\'en speed. The wave energy flux can be calculated as
\begin{equation}
{\bf F}={\bf F_{ac}}+{\bf F_{mag}}=p_1{\bf v}+{\bf B_1}\times({\bf v}\times {\bf B_0})/\mu_0,
\label{eq:F}
\end{equation}

\noindent where $p_1$ and ${\bf B_1}$ are the perturbations in the pressure and magnetic field, respectively, ${\bf v}$ is the velocity, ${\bf B_0}$ is the equilibirum magnetic field, and $\mu_0$ is the magnetic permeability. 

Figures \ref{fig:flux5}, \ref{fig:flux10}, and \ref{fig:flux15} present the acoustic \emph{(top panels)} and magnetic \emph{(bottom panels)} energy fluxes obtained from Eq. (\ref{eq:F}). They clearly show that most of the energy introduced by the driver keeps below the layer $c_S=v_A$ (white horizontal line), since the low frequency (3.3 mHz band) slow acoustic mode can only propagate horizontally, and the fast magnetic mode above this layer returns towards the photosphere and generates new acoustic and magnetic flux after its transformation. Only the slow acoustic mode with frequencies above the cut-off frequency that propagates upwards along field lines, visible in top panel of Figs. $\ref{fig:flux5}-\ref{fig:flux15}$ at a radial distance around the location of the driver and $z$ between 0.5 and 1 Mm, supplies energy to the chromosphere, since the energy of the Alfv\'en mode is negligible, as we will show later. Most of the magnetic flux above the layer $c_S=v_A$ corresponds to the fast magnetic waves which are being reflected.

\begin{figure}
\includegraphics[width=80mm]{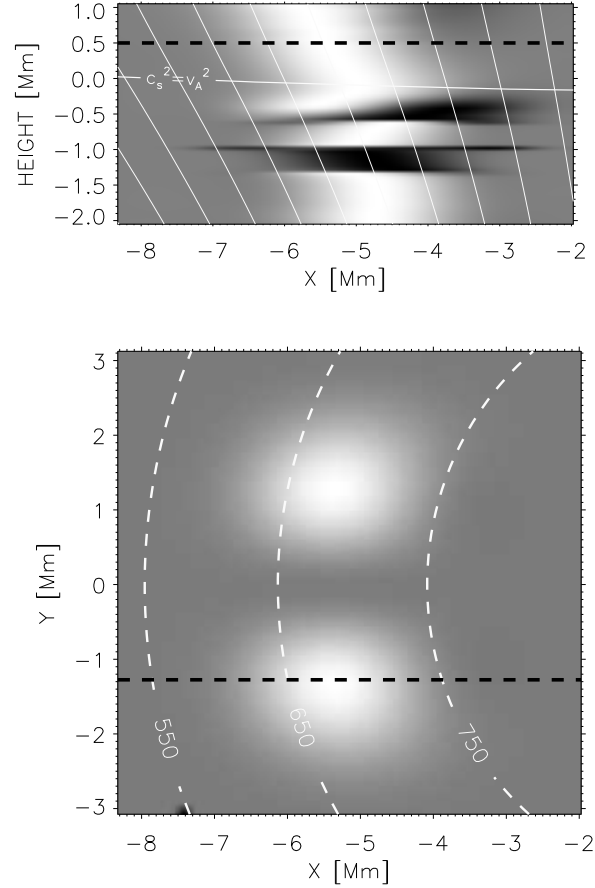}
\caption{Longitudinal magnetic flux of the Alfv\'en mode at $t=330$ s. \emph{Top:} Vertical cut in the plane $y=-1.3$ Mm, normalized at every height. Vertical white lines are magnetic field lines and horizontal white line is the layer where $c_S^2=v_A^2$. \emph{Bottom:} Horizontal cut in the plane $z=0.5$ Mm. Thin dashed lines are contours of equal magnetic field, with its value indicated in Gauss. In both panels, black dashed lines mark the location of the other plot.}
\label{fig:alfven}
\end{figure}

The Alfv\'en mode is visible in Fig. \ref{fig:alfven}. It shows the longitudinal magnetic energy flux calculated from Eq. (\ref{eq:F}) when ${\bf v}$ and ${\bf B_1}$ are projections along the polarization direction of the Alfv\'en wave \citep{Cally+Goossens2008}, for the driver located at 5 Mm from the axis of the sunspot. The top panel is a cut in the plane $x-z$ at $y=-1.3$ Mm, and it is normalized to its maximum at every height. The bottom panel is a horizontal cut in the plane $x-y$ at $z=0.5$ Mm. The Alfv\'en direction projections from \citet{Cally+Goossens2008} are valid only in asymptotic case $\beta \ll 1$. It means that in the top panel we only can indentify the Alfv\'en mode energy flux in the layers above the line $c_s^2=v_A^2$, and the flux below this height is not meaningfull. The Alfv\'en wave propagates upwards along field lines. The transformation from the fast acoustic mode to the Alfv\'en wave is null at the plane $y=0$ Mm, where the driver is located, but out of this plane there is energy flux in the Alfv\'en mode, as can be seen in the bottom panel. However, for the inclination of the magnetic field and the direction of propagation of the wave in these simulations, the trasformation to the Alfv\'en mode is very ineffective, and the magnetic flux of this wave at the height $z=0.5$ Mm is 40 times lower than the acoustic flux of the slow mode.

\section{Conclusions}

\begin{itemize}

 \item The mechanism that produces the change in frequency of oscillations in the umbra from the photosphere to the chromosphere is the linear propagation of waves with 3 minute power which come directly from the photosphere.

 \item The dominant wave frequency in the chromosphere decreases with the radial distance, due to the reduction of the cut-off frequency far from the axis.

 \item Only waves in the 5-6 mHz frequency band can provide energy to the chromosphere through slow acoustic mode, since waves with lower frequencies form evanescent waves and do not carry energy, and the fast magnetic mode is reflected back towards the photosphere. 

 \item The conversion to the Alfv\'en mode is effective where the magnetic field has some inclination, but its energy flux is small compared to the acoustic energy flux.

 \item Most of the energy is confined to the region below the height where the sound and Alfv\'en velocity are similar.

\end{itemize}

\end{document}